\documentclass[graybox]{svmult}

\usepackage{mathptmx}       
\usepackage{helvet}         
\usepackage{courier}        
\usepackage{type1cm}        
\usepackage{makeidx}         
\usepackage{graphicx}        
\usepackage{multicol}        
\usepackage[bottom]{footmisc}

\usepackage{amsfonts,amssymb,amsmath,bm,bbm}
\usepackage{graphicx,epsfig,color}
\usepackage{url}
\usepackage{comment}

\makeindex

\begin{document}

\title{Graphical Models and Belief Propagation-hierarchy for Optimal Physics-Constrained Network Flows}
\titlerunning{Graphical Models for Optimal Flows}

\author{Michael Chertkov, Sidhant Misra, Marc Vuffray, Dvijotham Krishnamurty, and Pascal Van Hentenryck}
\authorrunning{M. Chertkov,  S. Misra, M. Vuffray, D. Krishnamurthy, and P. Van Hentenryck}
\institute{
Michael Chertkov \at Theoretical Division, T-4 \& CNLS,
Los Alamos National Laboratory
Los Alamos, NM 87545, USA and
Energy System Center, Skoltech, Moscow, 143026, Russia, \email{chertkov@lanl.gov}
\and Sidhant Misra \at Theoretical Division, T-5,
Los Alamos National Laboratory
Los Alamos, NM 87545, USA, \email{sidhant@lanl.gov}
\and Marc Vuffray \at Theoretical Division, T-4,
Los Alamos National Laboratory
Los Alamos, NM 87545, USA, \email{sidhant@lanl.gov}
\and
Krishnamurthy Dvijotham \at Pacific Northwest National Laboratory, PO Box 999, Richland, WA 99352, USA \email{krishnamurthy.dvijotham@pnnl.gov}
\and Pascal Van Hentenryck \at University of Michigan, Department of Industrial \& Operations Engineering
Ann Arbor, MI 48109, USA,
\email{pvanhent@umich.edu}
}
\maketitle

\abstract{In this manuscript we review new ideas and first results on application of the Graphical Models approach, originated from Statistical Physics, Information Theory, Computer Science and Machine Learning, to optimization problems of network flow type with additional constraints related to the physics of the flow. We illustrate the general concepts on a number of enabling examples from power system and natural gas transmission (continental scale) and distribution (district scale) systems.}

\section{Introductory remarks}

In this chapter we discuss optimization problems which appears naturally in the classical settings describing flows over networks constrained by the physical nature of the flows which appear in the context of electric power systems, see e.g. \cite{01GS,08MBB}, and  natural gas application, see e.g. \cite{10Bor} and references there in. Other examples of physical flows where similar optimization problem arise include pipe-flow systems, such as district heating \cite{86Zin,DistrictHeatingWiki} and water \cite{16Rau}, as well as traffic systems \cite{99Lie}. We aim to show that the network flow optimization problem can be stated naturally in terms of the so-called Graphical Models (GM). In general, GMs for optimization and inference are wide spread in statistical disciplines such as Applied Probability, Machine Learning and Artificial Intelligence \cite{88Pea,96Jen,99CDLS,06Bis,09KF,12Mur}, Information Theory \cite{08RU} and Statistical Physics \cite{09MM}.

Main benefit of adopting GM methodology to the physics-constrained network flows is in modularity and flexibility of the approach -- any new constraints, set of new variables, and any modification of the optimization objective can be incorporated in the GM formulation with an ease. Besides, if all (or at least majority of) constraints and modifications are factorized,  i.e. can be stated in terms of a small subset of variables, underlying GM optimization or GM statistical inference problems can be solved exactly or approximately with the help of an emerging set of techniques, algorithms and computational approaches coined collectively  Belief Propagation (BP), see e.g. an important original paper \cite{05YFW} and recent reviews \cite{09MM,08RU,08WJ}. It is also important to emphasize that an additional benefit of the GM formulation is in its principal readiness for generalization. Even though we limit our discussion to application of the GM and BP framework to deterministic optimizations, many probabilistic and/or mixed generalizations (largely not discussed in this paper) fit very naturally in this universal framework as well.

We will focus on optimization problems associated with Physics-Constrained Newtork Flow (PCNF) problems. Structure of the networks will obviously be inherited in the GM formulation, however indirectly - through graph- and variable- transformations and modifications. Specifically,  next Section \ref{sec:formulations} is devoted solely to stating a number of exemplary energy system formulations in GM terms. Thus, in Section \ref{subsec:DONF} and Section \ref{subsec:GNF}  we consider dissipation optimal and respectively general physics-constrained network flow problems. In particular, Section \ref{subsec:GNF} includes discussion of power flow problems in both power-voltage, Section \ref{subsubsec:AC-power}, and current-voltage, Section \ref{subsubsec:AC-current}, formats, as well as discussion of the gas flow formulation in Section \ref{subsubsec:gas} and general k-component physics-constrained network flow problem in Section \ref{subsubsec:k}. Section \ref{subsec:opt_PCNF} describes problems of the next level of complexity -- these including optimization over resources. In particular, general optimal physics-controlled network flow problem is discussed in Section \ref{subsubsec:opt_gen} and more specific cases of optimal flows, involving optimal power flow (in both power-flow and current-voltage formulations) and gas flows are discussed in Sections \ref{subsubsec:opt-AC},\ref{subsubsec:opt-IV},\ref{subsubsec:gas}, respectively. Section \ref{subsec:feasibility} introduces a number of feasibility problems, all stated as special kinds of optimizations. Here we discuss the so-called instanton,
Section \ref{subsubsec:instanton}, containment Section \ref{subsubsec:containment}, and state estimation, Section \ref{subsubsec:state_estimation}, formulation. The long introductory section concludes with a discussion in Section \ref{subsec:aggregators} of an exemplary (and even more) complex optimization involving split of resources between participants/aggregators.

In Section \ref{sec:GM} we describe how any of the aforementioned PCNF and optimal PCNF problems can be re-stated in the universal Graphical Model format.

Then, in Section \ref{sec:LP-BP}, we take advantage of the factorized form of the PCNF GM and illustrate how BP methodology can be used to solve the optimization problems exactly and/or approximately. Specifically, in Section \ref{subsec:NP_to_LP_exact} we restate the optimization (Maximum Likelihood) GM problem as a Linear Programming (LP) in the space of beliefs (proxies for probabilities). The resulting LP is generally difficult as working with all variables in a combination. We take advantage of the GM factorization  and  introduce in Section \ref{subsec:LP-BP_relax} the so-called Linear Programming - Belief Propagation (LP-BP) relaxation, providing a provable lower bound for the optimal. Finally, in Section \ref{subsec:LP-BP_int} we construct  a tractable relaxation of LP-BP based on an interval partitioning of the underlying space.

Section \ref{sec:hierarchies} discuss hierarchies which allow to generalize, and thus improve LP-BP. The so-called LP-BP hierarchies, related to earlier papers on the subject  \cite{02Wai,08Joh,10Son} are discussed in Section \ref{subsec:LP-BP-hierarchy}. Then, relation between the LP-BP hierarchies and classic LP-based Sherali-Adams \cite{90SA} and Semi-Definite-Programming based Lasserre  hierarchies \cite{91LS,01Las,03Par,10Las} are discussed in Section \ref{subsec:hierarchy-relations}.

Section \ref{sec:tree} discuss the special case of  a GM defined over a tree (graph without loops). In this case LP-BP is exac, equivalent to the so-called Dynamic Programming approach, and as such it provides a distributed alternative to the global optimization through a sequence of graph-element-local optimizations. However, even in the tree case the exact LP-BP and/or DP are not tractable for GM stated in terms of physical variables, such as flows, voltages and/or pressures, drawn from a continuous set. Following, \cite{Dvijotham2016} we discuss here how the problem can be resolved with a proper interval-partitioning (discretization).

We conclude the manuscript presenting summary and discussing path forward in Section \ref{sec:conclusions}.

\section{Problems of Interest: Formulations}
\label{sec:formulations}

In this Section we formulate a number of physics-constrained network flow problems which we will then attempt to analyze and solve with the help of Graphical Model (GM)/Belief Propagation (BP) approaches/techniques in the following Sections.

\subsection{Dissipation-Optimal Network Flow}
\label{subsec:DONF}

We start introducing/discussing Network Flows constrained by a minimum dissipation principle, i.e. one which can be expressed as an unconstrained optimization/minimization of an energy function (potential).

Consider a static flow of a commodity over an undirected graph, ${\cal G}=({\cal V},{\cal E})$ described through the following network flow equations
\begin{eqnarray}
i\in{\cal V}:\quad q_i=\sum_{j:(i,j)\in {\cal E}} \phi_{ij}, \label{flow_eq}
\end{eqnarray}
where $q_i$ stand for injection,  $q_i>0$, or consumption, $q_i<0$, of the flow at the node $i$ and $\phi_{ij}=-\phi_{ji}$ stands for the value of the flow through the directed edge $(i,j)$ -- in the direction from $i$ to $j$ \footnote{In the following we will use notation $\{i,j\}$ for the undirected graph and $(i,j)$ for the respective directed graph. When the meaning is clear we slightly abuse notations denoting by ${\cal E}$ both the set of undirected and directed edges.}. We consider a balanced network, $\sum_{i\in{\cal V}} q_i=0$.

We constraint the flow requiring that the minimum dissipation principle is obeyed
\begin{eqnarray}
\left. \min_\phi \sum_{\{i,j\}\in{\cal E}} E_{ij}(\phi_{ij})\right|_{\mbox{Eq.~(\ref{flow_eq})}},
\label{convex_cost_opt}
\end{eqnarray}
where $\phi\doteq(\phi_{ij}=-\phi_{ji}|\{i,j\}\in {\cal E})$, and $E_{ij}(x)$ are local (energy) functions of their arguments for all $\{i,j\}\in{\cal E}$. The local energy functions  $E_{ij}(x)$ are required to be convex at least on a restricted domain.
We call the sum of local energy functions $E(\phi)=\sum_{\{i,j\}\in{\cal E}} E_{ij}(\phi_{ij})$ the global energy function or simply the energy function. Versions of this problem appear in the context of the feasibility analysis of the dissipative network flows, that is flows whose redistribution over the network is constrained by potentials,  e.g. voltages or pressures in the context of resistive electric networks and gas flow networks, respectively \cite{2015arXivMonotoneDJ,
2015arXivMaxThroughput,2015CDCMonotonicity}. Note,  that the formulation (\ref{convex_cost_opt}) can also be supplemented by additional flow or potential constraints.

Requiring Karush–-Kuhn–-Tucker (KKT) stationary point conditions on the optimization problem stated in Eq.~(\ref{convex_cost_opt}) leads to the following set of equations
\begin{eqnarray}
\forall \{i,j\}\in{\cal E}:\quad E'_{ij}(\phi_{ij})= \lambda_i-\lambda_j,
\label{KKT}
\end{eqnarray}
where $\lambda_i$ is a Lagrangian multiplier corresponding to the i's equation (\ref{flow_eq}).  The problem becomes fully defined by the pair of Eqs.~(\ref{flow_eq},\ref{KKT}), which can also be re-stated solely in terms of the $\lambda$-variables
\begin{eqnarray}
i\in{\cal V}:\quad q_i=\sum_{j:\{i,j\}\in {\cal E}} \left(E'_{ij}\right)^{-1}(\lambda_i-\lambda_j). \label{flow_eq_lambda}
\end{eqnarray}

\subsection{General Physics-Constrained Network Flows}
\label{subsec:GNF}

We call ``unconstrained" a network flow for which only conservation of flow(s), described by Eq.~(\ref{flow_eq}), is enforced. Contrariwise we call ``Physics-constrained" a network flow that in addition to flow conservation also enforces constrains relating line flows bounding nodal values of a physics potential, e.g. voltages, pressures, etc.

A particular example of the physics-constrained network flow was discussed above in Section \ref{subsec:DONF}. However, this example is special as it represented network flows as a gradient of a scalar energy function.
Aiming to discuss the general case, where a physics-constrained network flow problem cannot be stated as one following from minimization of a scalar energy function,  we find it useful to start below with an example of the AC electric power flow and then proceed to discussing an abstract general case.

\subsubsection{AC Network Flow: power-voltage formulation}
\label{subsubsec:AC-power}

AC power flows,  where one accounts for both inductance and resistivity of lines, is one of the main example of the physics-constrained network flow. We thus start discussing the AC flow described in terms of the set of algebraic equations over the graph, ${\cal G}$:
\begin{eqnarray}
&& \forall i\in{\cal V}:\quad P_i=\sum_{j:(i,j)\in{\cal E}} \phi_{ij},\label{AC_PF_1}\\
&& \forall (i,j)\in{\cal E}:\quad \phi_{ij}=V_i\left(\frac{V_i-V_j}{z_{ij}}\right)^* \label{AC_PF_2}
\end{eqnarray}
where all the characteristics take values over complex numbers, $V_i$ is the complex voltage potential, $\phi_{ij}$ is the complex power leaving node $i$ in the direction to node $j$, $P_i$ are complex injections/consumptions at the nodes, $z_{ij}=z_{ji}$ is the complex impedance of the line $\{i,j\}$ (assumed known); and $y^*$ stands for the complex conjugate of $y$. One formulation/problem of interest is: given $P_i$ at all node but one ($i=0$ called slack bus),  and fixing the voltage at the slack bus, e.g. $V_0=1$, to find $V_i$ at $\forall i\in{\cal V}\setminus 0$.

In general the full formulation Eqs.~(\ref{AC_PF_1},\ref{AC_PF_2}) cannot be represented as a gradient of a scalar function of voltages. However, such representation is possible in the special cases, when one either neglect resistance in lines, in comparison with inductance, or when all the lines of the system are characterized by a constance inductance-to-resistance ratio.

Power systems may include transformers of different type, e.g. standard voltage transformers or phase transformers. These devices can be described as nodes of degree two. For instance consider multiplicative transformations
\begin{eqnarray}
\forall i\in{\cal V}_T\subset{\cal V}:\quad V_{i;in}=\alpha_i V_{i;out}
\label{transf}
\end{eqnarray}
that are characterized by the complex transformation coefficient,  $\alpha_i$. For phase-transformers $|\alpha_i|=1$ if losses of actual and reactive power at the transformer are ignored. Other type of transformations, e.g. additive or generally nonlinear, can be easily incorporated in the model. Even though already installed transformers are typically not used for the real-time control in practical transmission systems, the newly installed solid-state transformers are capable of fast and efficient response and thus they can actually be used for real-time (even seconds-scale) controls. High-Voltage-Direct-Current (HVDC) links are new installations which can also be incorporated into the PF description. The HVDC can be modeled as a pair of points, or multiple points for multi-terminal HVDC, with a zero-net injected/withdrawed active and reactive power if we assuming that the devices are lossless.

Finally,  let us mention that lines can be modeled in a more accurate way via the so-called $\pi$-model. We will not describe it here in details, only mentioning that this modeling fits naturally the general graph-description of the systems as it simply requires introducing two auxiliary nodes at the two ends of the line connected through capacitors to the ground.

\subsubsection{AC Network Flow: current-voltage formulation}
\label{subsubsec:AC-current}

The PF Eqs.~(\ref{AC_PF_1},\ref{AC_PF_2}) can also be restated in terms of the linear Kirchoff law relations between currents and voltages
\begin{eqnarray}
&& \forall i\in{\cal V}:\quad I_i=\sum_{j:(i,j)\in{\cal E}} J_{ij},\label{AC_PF_currents_1}\\
&& \forall (i,j)\in{\cal E}:\quad J_{ij}=\frac{V_i-V_j}{z_{ij}}, \label{AC_PF_currents_2}
\end{eqnarray}
where $P_i=V_i I_i^*$ and $\phi_{ij}=V_i J_{ij}^*$. When focus is on resolving the PF problem -- given nodal consumptions and productions of power, $P$, one aims to find voltages, $V$, and power flows, $\phi$, over lines 
-- the nonlinear PF formulation due to Eqs.~(\ref{AC_PF_1},\ref{AC_PF_2}) is primal.  However, as argued below the Kirchoff original formulation (\ref{AC_PF_currents_1},\ref{AC_PF_currents_2}) may offer some additional computational advantages for posing and solving optimal problems where the power production and consumption is an optimization variable that is not fixed to a pre-defined value.

\subsubsection{Gas Flows}
\label{subsubsec:gas}

Balanced Gas Flows (GF) satisfy the following set of algebraic equations
\begin{eqnarray}
&& \forall i\in{\cal V}:\quad q_i=\sum_{j:(i,j)\in{\cal E}} \phi_{ij},\label{GF_1}\\
&& \forall (i,j)\in{\cal E}:\quad \phi_{ij}=\gamma_{ij}\frac{|\pi_{i} - \pi_{j} + b_{ij}|^{3/2}}{\pi_{i} - \pi_{j} + b_{ij}}, \label{GF_2}\\
&& \forall i\in {\cal V}_c\subset{\cal V}:\quad \pi_{i;out}=\alpha_i \pi_{i;in},\label{GF_3}
\end{eqnarray}
where $\pi_i\geq 0$ is the squared pressure at  node $i$; $\gamma_{ij}$ is a constant characterizing the line or pipe $\{i,j\}$ which depends on diameter of the pipe, friction coefficient, the type of gas used, etc; $b_{ij}$ is a coefficient of an additive compression at the pipe $\{i,j\}$ and $\alpha_i$ is a coefficient of a multiplicative compression at the compressor node, $\i\in{\cal V}_c$, which is normally a node of degree two, and then $\pi_{i;in}$ and $\pi_{i;out}$ stand for squared pressures at both sides of the node. Both types of compressors can be present at a line but not simultaneously, depending on possible operational strategies. Like in the PF case,  it is also convenient to assume existing of a slack bus, which also reflects a practical situation. The slack bus is a special node, $i=0$, where the pressure is maintained constant providing a source for the global balance of the gas flow.

\subsubsection{General Physics-Constrained Network Flows}
\label{subsubsec:k}

A general $K$-component Physics-Constrained Network Flow (PCNF) problem becomes
\begin{eqnarray}
&& \forall k=1,\cdots,K,\ i\in{\cal V}:\quad q_i^{(k)}=\sum_{j:(i,j)\in {\cal E}} \phi_{ij}^{(k)}, \label{flow_eq_repl}\\
&& \forall k=1,\cdots,K,\ (i,j)\in{\cal E}:\quad \phi_{ij}^{(k)}= f_{ij}^{(k)}(\pi_i,\pi_j),
\label{flow_general}
\end{eqnarray}
where $\pi_i\doteq(\pi_i^{(k)}|k=1,\cdots,K)$. Nodal transformers/compressors can be readily included into the model
\begin{eqnarray}
 i\in{\cal V}_t\subset{\cal V}:\quad \pi_{i;out}=T_i(\pi_{i;in}),
\label{general-transformer-compressor}
\end{eqnarray}
where $T_i(\cdot)$ can be a general nonlinear transformation and $\pi_i=(\pi_i^{(k)}|k=1,\cdots,K)$.

\subsection{Optimal Physics-Constrained Network Flow Problems}
\label{subsec:opt_PCNF}

The Optimal Physics-Constrained Network Flow (OPCF) problems aim to find an optimum over a set of control/optimization parameters that enter into the physics-constrained network flow description.

We will first present a definition of OPCFs in a rather general setting, and later illustrate this problem with some examples.

\subsubsection{General Case}
\label{subsubsec:opt_gen}
In the most general case we want to solve the following optimization problem
\begin{eqnarray}
\left.\min_{q,\pi,\phi,\{T\}} \left(\sum_{i\in{\cal V}} C_i(q_i)+\sum_{i\in{\cal V}_t} C^{(t)}_i\{T_i\}\right)\right|_{\begin{array}{cc}
\mbox{Eqs.~(\ref{flow_eq_repl},\ref{flow_general},\ref{general-transformer-compressor})} & \\
\pi_i\in \Pi_i &\forall i\in{\cal V}\\
\phi_{ij}\in \Psi_{ij}&\forall (i,j)\in{\cal E}
\end{array}}
\label{opt_thr_gen}
\end{eqnarray}
where $\Pi_i$ and $\Psi_{ij}$ describe the domains of allowed values for node-potentials and edge-flows, respectively.

\subsubsection{Optimal Power Flow: power-voltage formulation}
\label{subsubsec:opt-AC}

Standard Optimal Power Flows (OPF) over transmission systems are stated as follows, see e.g. \cite{13Bie,15Bie} and references therein,
\begin{eqnarray}
\min_{P,\Phi,V} \left.\sum_{i\in{\cal V}} C_i(P_i)\right|_{
\begin{array}{cc}
\mbox{Eqs.~(\ref{AC_PF_1},\ref{AC_PF_2})}&\\
V_i\in U_i &\forall i\in{\cal V}\setminus 0\\
\phi_{ij}\in \Psi_{ij}&\forall (i,j)\in{\cal E}
\end{array}}
\label{OPF}
\end{eqnarray}
where $V_0=1$, $C_i(P_i)$ is the cost function that is potentially nonlinear and site-dependent and $U_i$, $\Psi_{ij}$ are domains of allowed values for site-voltage  and line-flows, respectively. There are multiple other extensions, generalizations, e.g. accounting for investment and planning of new devices, such as FACTS, HVDC and transformer devices, see e.g.  \cite{14FBCa,14FBCb,16FGBBC}.

The OPF problem (\ref{OPF}) gets simpler in the case of the distribution grid where the graph is a tree. Then, voltage is fixed at the head of the tree, $i=0$, considered as a slack bus, while all other nodes of the system are modeled in the static setting as $(p,q)$ nodes, where $p_i$ is an accumulated consumption and Photo-Voltaic (PV) generation at the node, $i$, and $q_i$ is the reactive power consumed/produced at the node. PV power is injected to the grid through inverters,  which have a capability to adjust reactive power.  This degree of freedom can be used to achieve various objectives, e.g. to minimize (active) power losses in lines subject to voltages to stay within pre-defined safety limits. An exemplary distribution grid OPF is
\begin{align}
 \min_{q,V} \quad & \sum_{\{i,j\}\in{\cal E}}\frac{|V_i-V_j|^2}{r_{ij}^2+x_{ij}^2}r_{ij}, \label{OPF_dist}\\
 \mbox{ s.t. } \quad &
\begin{array}{cc}
p_i+{\it i} q_i=V_i\sum_{j:\{i,j\}\in{\cal E}}\left(\frac{V_i-V_j}{z_{ij}}\right)^*,& \forall i\in{\cal V}\setminus 0  \\
V_i\in U_i \ \& \ q_i\in {\cal Q}_i &\forall i\in{\cal V}
\end{array}
\label{OPF_dist_constr}
\end{align}
where $q\doteq (q_i|i\in{\cal V}\setminus 0)$, $V=(V_i\in \mathbb{C}|i\in{\cal V}\setminus 0)$ are variable vectors of reactive injections/consumptions and voltages, and the vector of active injection/consumption, ${\cal Q}_i$ describes the allowed range of the nodal reactive power adjustment; $p=(p_i\in \mathbb{R}|i\in{\cal V}\setminus 0)$ is assumed fixed; and $U_0=\{1\}$,  i.e. voltage at the head of the line is constrained.  Notice, that given that the underlying graph is a tree the PF equations can be rewritten in the so-called Baran-Wu representation \cite{89BW}, stated in terms of both active and reactive power flows flowing through the line segments, and voltages at the nodes. Note that the Baran-Wu representation also applies to loopy networks, however in the loopy case the related system of equations is incomplete i.e. underdefined.

\subsubsection{Optimal Power Flow: current-voltage formulation}
\label{subsubsec:opt-IV}

Assume that all nodes of the network have some kind of flexibility in terms of injection/consumption, i.e. $I_i$ is not fixed but is allowed to be drawn from range $\Xi_i$ that is potentially node specific. Then one poses the following current-voltage version of the OPF formulation
\begin{eqnarray}
\min_{I,J,V} \left.\sum_{i\in{\cal V}} C_i(V_i I_i^*)\right|_{
\begin{array}{cc}
\mbox{Eqs.~(\ref{AC_PF_currents_1},\ref{AC_PF_currents_2})}&\\
V_i\in U_i \ \&\  I_i\in \Xi_i &\forall i\in{\cal V}\setminus 0\\
(V_i-V_j)J_{ij}^*\in \Psi_{ij}&\forall (i,j)\in{\cal E}
\end{array}}
\label{OPF_current}
\end{eqnarray}

\subsubsection{Optimal Gas Flow}
\label{subsubsec:opt-gas}

A rather general version of the optimum gas flow problem is
\begin{eqnarray}
\left.\min_{q,p,\alpha} \left(\sum_{i\in{\cal V}} C_i(q_i)+\sum_{i\in{\cal V}_\alpha} C_i(\alpha_i)\right)\right|_{\begin{array}{cc}
\mbox{Eqs.~(\ref{GF_1},\ref{GF_2},\ref{GF_3})} & \\
\pi_i\in \Pi_i &\forall i\in{\cal V}\\
\phi_{ij}\in \Psi_{ij}&\forall (i,j)\in{\cal E}
\end{array}}
\label{opt_gas_flow}
\end{eqnarray}
where ${\cal V}_\alpha$ is the set of the multiplicative compressor nodes, $\alpha=(\alpha_i|i\in{\cal V}_\alpha)$ is the vector of compression; and the two contributions to the objective balance deviation of the consumption/injection of gas from the nominal value across the system with the cost of compression. See \cite{68Wl,osiadacz1987simulation,00WS,00WRBS,12BNV,15MFBBCP,2015CDCMonotonicity} for additional details.

\subsection{Feasibility as an Optimal Physics Constrained Network Flow (PCNF) problem}
\label{subsec:feasibility}

Problems discussed below can all be understood as network feasibility problems of special types which focus on describing or characterizing domains of feasibility of operations.  Suggesting good algorithms for efficient and accurate solutions of these problems will allow to monitor state of the system not as one particular configuration
but as a succinct characterization of the domains with good or bad properties. Thus the problem can also be described as guiding, building or focusing on an "extended state evaluations or characterizations".

\subsubsection{Instanton as an Optimal PCNF problem}
\label{subsubsec:instanton}

An instanton is a special network flow state, $(\phi,\pi)_{inst}$, which is defined as the most probable failure state. Consider, for example stochastic injections/consumptions, $q$, drawn from an exogenously known probability distribution, ${\cal P}(q)$. The probability is viewed as a distance measure, $D(q;q_0)=\log({\cal P}(q_0)/{\cal P}(q))$, from the most probable configuration of the injection/consumption, $q_0$. In many practicle cases $D(q;q_0)$ shows nice properties, e.g. $D(q;q_0)$ is a convex function of $q$. A state, $(\phi,\pi)$, is considered faulty if it is on the boundary, $(\phi,\pi)\in B_{\text{safe}}$, of the domain of the safe operation. Therefore the instanton problem in the case of a general PCNF flow is a solution of the following optimization problem
\begin{eqnarray}
\left.\min_{(\phi,\pi)} D(q;q_0)\right|_{
\begin{array}{c}
\mbox{Eqs.~}(\ref{flow_eq_repl},\ref{flow_general})\\
(\phi,\pi)\in B_{\text{safe}}
\end{array}
}
\label{instanton}
\end{eqnarray}
Description of the boundary domain, $B_{\text{safe}}$, will depend on what is considered ``safe".  Two examples of interest are boundaries of (a) union of the box constraints on line flows; (b) boundary of the PCNF feasibility,  i.e. the domain where the determinant of the respective Jacobian is zero. Considering boundary of the intersection of the two exemplary domains is also of interest. See \cite{11CPS,11MSPB,15KHCBB} for additional details.

\subsubsection{Containment as an Optimal PCNF problem}
\label{subsubsec:containment}

Suppose we identify ``desirable properties" in a space of operational parameters, $(\phi,\pi)$, such as voltages, pressures, power flows, etc. The special features of the ``desirable" domain, ${\cal D}_{\text{des}}$, may allow simpler characterization of the domain. The examples are convexity of an underlying energy function, monotonicity of an underlying operator, piece-wise monotonicity in the response of the system, or simply existence of a solution.  Description of ${\cal D}_{\text{des}}$ may be algebraically nontrivial, e.g. stated as a non-negativity of a matrix, positivity of the largest eigen-value of a matrix, or positivity of all components of a matrix. On the other hand we may have an alternative description of a ``safety" domain, ${\cal D}_{\text{safe}}$,  in space of operational parameters.  For example,  we may want flows over lines not to exceed respective thresholds, voltages or pressures to be within bounds, etc.
Description of both the ``desirable property" and ``safety" domains may allow some additional degrees of freedom which will be changing shapes of the domain, e.g. making them larger or smaller, fitting a certain shape, etc. For example, we may consider the ``safety" domain to depend on a re-scaling volume factor, $V$: ${\cal D}_{\text{safe}}(V)$.  The containment problem becomes, to optimize the additional degrees of freedom in the description of both the ``desirable" domain and/or the ``safety" domain, e.g. $V$, so that the latter would be contained within the former. Formally, the containment problem is stated as the following optimization problem
\begin{eqnarray}
\left.\min_{(\phi,\pi); V} V\right|_{
\begin{array}{c}
\mbox{Eqs.~}(\ref{flow_eq_repl},\ref{flow_general})\\
{\cal D}_{\text{safe}}(V)\subseteq {\cal D}_{\text{des}}
\end{array}
}
\label{containment}
\end{eqnarray}
See \cite{15DLCa,15DK,15DLCb,15DCL} for additional details.

\subsubsection{State Estimation as an Optimal PCNF problem}
\label{subsubsec:state_estimation}

Here we discuss a data driven state estimation problem: given deterministic or probabilistic measurements, describe a state or domain of states that are most consistent with the data. For example, consider the observational data, e.g. measured by PMU in the case of power systems, to be a subset of line flows, $\phi_{\text{d}}=(\phi_{ij;\text{d}}^{(k)}|(i,j)\in{\cal E}_{\text{d}}\subseteq {\cal E}; \forall\ k-1,\cdots,K)$, and potentials, $\pi_{\text{d}}(\pi_{i;\text{d}}^{(k)}|i\in{\cal V}_{\text{d}}\subseteq {\cal V}; \forall\ k-1,\cdots,K))$ measured respectively at ${\cal V}_{\text{d}}$ and ${\cal E}_{\text{d}}$, respectively. Then an exemplary data-most consistent state estimation problem can be found by solving the following optimization problem
\begin{align}
\min_{q;(\phi,\pi)} \quad &\sum_{k=1,\cdots,K; i\in{\cal V}}\parallel q_i^{(k)}-\sum_{j:(i,j)\in {\cal E}} \phi_{ij}^{(k)}\parallel
\label{state_estimation}\\
\mbox{ s.t. } \quad &
\begin{array}{c}
\forall i\in {\cal V}_{\text{d}},\ \forall k=1,\cdots,K:\ \pi_i^{(k)}=\pi_{i;\text{d}}^{(k)}\\
\forall (i,j)\in {\cal E}_{\text{d}},\ \forall k=1,\cdots,K:\ \phi_{ij}^{(k)}=\phi_{ij;\text{d}}^{(k)}\\
\mbox{Eqs.~}(\ref{flow_general})
\end{array}
\label{state_estimation_cond}
\end{align}

\subsection{Optimal Physics-Constrained Network Flows with Resources Split Between Aggregators}
\label{subsec:aggregators}

In some cases energy resources and energy consumption can be redistributed between a group of nodes. For example, Electric Vehicle (EV) aggregator may split its EV flit in two or more groups to be charged at distinct locations.  Similarly, mobile battery resources can be re-distributed by a battery aggregator between two or more nodes. This type of dependencies can be modeled by introducing additional pair-wise or high-order constraints on the nodal injection-consumptions.

For example, consider the following generalization of the distribution system OPF (\ref{OPF_dist},\ref{OPF_dist_constr}) allowing for flexibility of resources split between a number of aggregators
\begin{align}
 \min_{q,V,p_c} \quad & \sum_{\{i,j\}\in{\cal E}}\frac{|V_i-V_j|^2}{r_{ij}^2+x_{ij}^2}r_{ij}, \label{OPF_dist_ext}\\
 \mbox{ s.t. } \quad &
\begin{array}{cc}
p_i+{\it i} q_i=V_i\sum_{j:\{i,j\}\in{\cal E}}\left(\frac{V_i-V_j}{z_{ij}}\right)^*,& \forall i\in{\cal V}\setminus 0  \\
V_i\in U_i \ \& \ q_i\in {\cal Q}_i &\forall i\in{\cal V}\\
\underline{p}_\alpha\leq |p_i+p_j|\leq \overline{p}_\alpha &\forall i,j\sim \alpha\in {\cal A}
\end{array}
\label{OPF_dist_constr_ext}
\end{align}
where ${\cal A}$ stands for the list of the pair-wise aggregators,  and $i,j\sim \alpha$ indicates that the two distinct nodes $i$ and $j$ are under control of the same aggregator. Generalization to aggregators controlling more than two nodes is straightforward.

\section{Graphical Model formulation  for a Physics-Constrained Optimal Network Flow problem}
\label{sec:GM}

\begin{figure}
\centering
\includegraphics[width=0.95\textwidth,page=1]{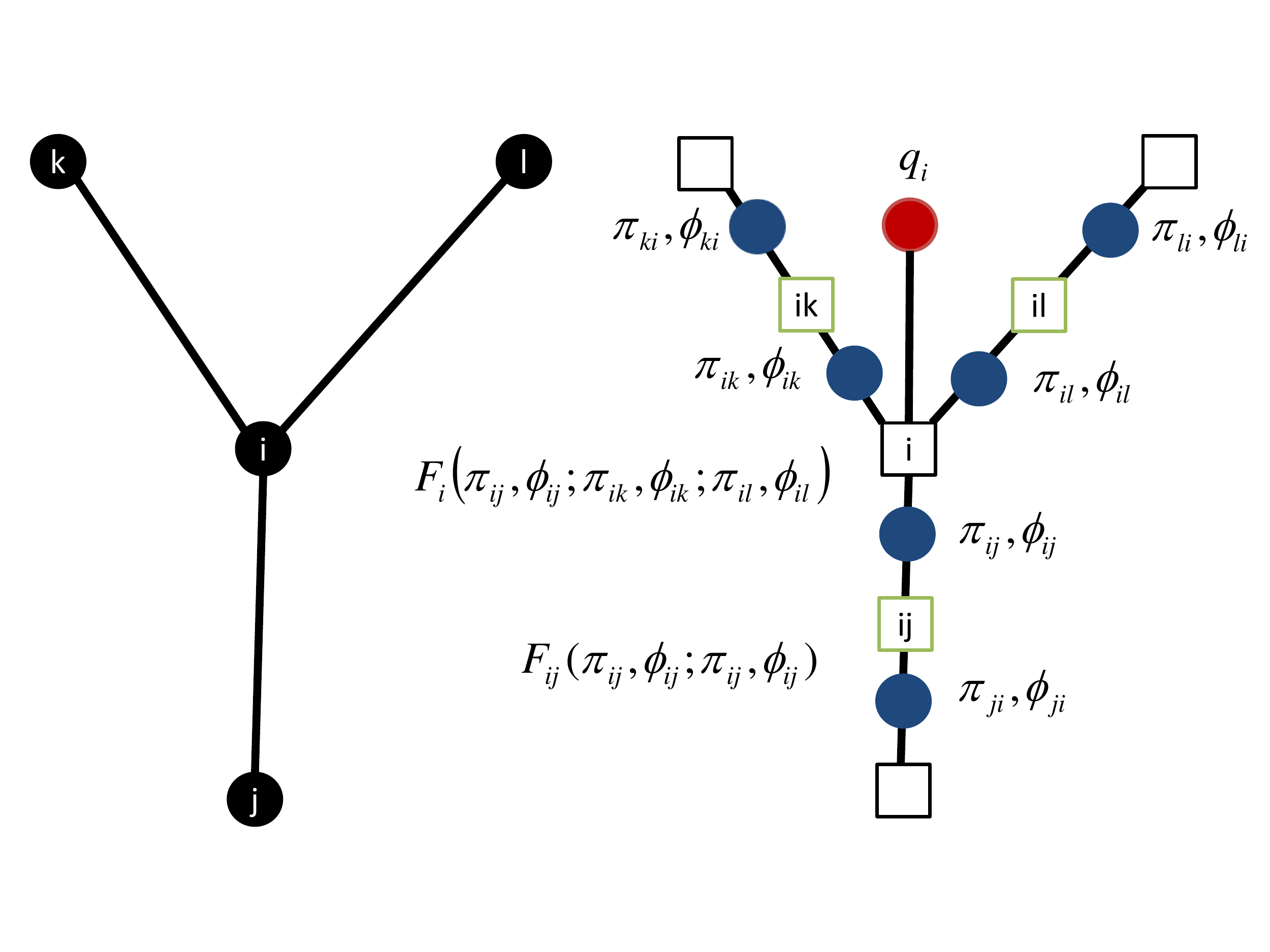}
\caption{Illustration of an element of the GM (\ref{GM1}), shown on the right, construction from respective element of the base physical network graph, shown on the left. Variable nodes of the GM are shown as the circles/nodes. Check/function nodes are shown as squares. Duplicated potentials, e.g. $\pi_{ij}$, and flows, e.g. $\phi_{ij}$, are associated with the blue circles/nodes and injections/consumptions, e.g. $q_i$ are associated with the red circles/nodes. Functions associated with the black and green checks/squares implement duplication and flow conservation, e.g. $F_i\left(q_i;
\pi_{\sim i};\phi_{\sim i}\right)$, and dissipative relation between for the flow drop over line as a function of potentials at the two ends of the line, e.g.
$F_{ij}\left(\pi_{ij};\phi_{ij};\pi_{ji}\right)$, defined in Eqs.~(\ref{Fi},\ref{Fij}) respectively.}
\label{graph-trans}
\end{figure}

In the general optimal PCNF formulation (\ref{opt_thr_gen}) the state/optimization vector, $s\doteq (\pi,\phi,q)$, or simply state, is represented by the vector of potentials,
$\pi\doteq(\pi_{ij}|(i,j)\in{\cal E})$, the vector of line flows, $\phi\doteq(\phi_{ij}|(i,j)\in{\cal E})$, where components of the latter are associated with the directed edges and thus assumed computed at the starting node of the edge, and the injection-consumption vector, $q=(q_i|i\in{\cal V})$.

Consider the following probabilistic version of the optimization problem (\ref{opt_thr_gen}) where the state $s$ is realized with a probability factorized according to the following distribution function
\begin{eqnarray}
&& {\cal P}(s)\sim \exp\left(-\beta\sum_{i\in{\cal V}\setminus 0} C_i(q_i)\right)
\prod_{i\in{\cal V}} F_i\left(q_i;\pi_{\sim i};\phi_{\sim i}\right)F_{ij}\left(\pi_{ij},\phi_{ij};\pi_{ji},\phi_{ji}\right),
\label{GM1}\\
&& \forall i\in{\cal V}\setminus 0:\quad F_i\left(q_i;\pi_{\sim i};\phi_{\sim i}\right)\doteq
\left\{\begin{array}{cc} 1, & (q_i;\pi_{\sim i};\phi_{\sim i})\in \Upsilon_i\\
0, & (q_i;\pi_{\sim i};\phi_{\sim i})\notin \Upsilon_i\end{array}\right. \label{Fi}\\
&& \Upsilon_i\doteq \left(\pi_{ik}=\pi_{ij}=\pi_{il},\ \& q_i=\phi_{ij}+\phi_{ik}+\phi_{il}\right),\label{Upsilon_i}\\
&& \forall (i,j)\in {\cal E}:\quad F_{ij}\left(\pi_{ij};\phi_{ij};\pi_{ji}\right)\doteq
\left\{\begin{array}{cc} 1, & (\pi_{ij};\phi_{ij};\pi_{ji})\in \Upsilon_{ij}\\
0, & (\pi_{ij};\phi_{ij};\pi_{ji})\notin\Upsilon_{ij}\end{array}\right. \label{Fij}\\
&& \Upsilon_{ij}\doteq \left(\phi_{ij}=f_{ij}(\pi_{ij},\pi_{ji}\right),\label{Upsilon_ij}
\end{eqnarray}
where  $\beta>0$ is an auxiliary parameter sometimes called inverse effective temperature; $\forall i\in{\cal V}:\quad \pi_{\sim i}\doteq (\pi_{ij}|(i,j)\in{\cal E})$ and $\phi_{\sim i}\doteq (\phi_{ij}|(i,j)\in{\cal E})$ are vectors of potentials and flows associated with a vertex;  $C_i(q_i)$ is the cost dependent on the consumption/injection, $q_i$, at the node $i$; and $\delta(x)$ is the characteristic function of the logical expression $x$: $\delta(x)$ is unity if $x$ is true, and it is zero otherwise. Let us assume that all the flow variables, i.e. all components of $\phi$ and $q$ vectors are drawn form  a finite alphabet, $\forall i\in{\cal V}:\quad q_i\in \Theta$ and $\forall (i,j)\in{\cal E}:\quad \phi_{ij}\in \Theta$. Let us also assume that the components of $\pi$ take values in a finite set, $\forall (i,j)\in {\cal E}:\quad \pi_{ij}\in\tilde{\Pi}$; and denote the resulting finite set for $s$ by $\Sigma$. Probability of the state $s\in\Sigma$ given by
Eq.~(\ref{GM1} can be understood as representing a Graphical Model constructed based on the physical network ${\cal G}=({\cal V},{\cal E})$, where the construction is illustrated in Fig.~\ref{graph-trans}.

Obviously the Maximum Likelihood (ML) configuration associated with ${\cal P}(s)$ from Eq.~(\ref{GM1}) corresponds to the solution of the Optimal Flow problem (\ref{opt_thr_gen}), which can also be restated in GM terms as follows
\begin{align}
E\doteq &\min_{s\in\Sigma} \quad \sum_{i\in{\cal V}\setminus 0} C_i(q_i)\label{ML_opt_flow1}\\
&\mbox{ s.t. } \quad
\begin{array}{cc}
\left(q_i;\pi_{\sim i};\phi_{\sim i}\right)\in\Upsilon_i & \forall i\in{\cal V}\setminus 0\\
\left(\pi_{ij};\phi_{ij};\pi_{ji}\right)\in\Upsilon_{ij} & \forall (i,j)\in {\cal E}.
\end{array}
\label{ML_opt_flow1_cond}
\end{align}

This Physics-Constrained Network Flow optimization problem is a special case of a general GM optimization problem
\begin{eqnarray}
&& \mbox{OPT:}\quad \min_{x\in \Sigma} \sum_{\alpha\in \bar{\cal V}_f} f_\alpha (x_\alpha)
\label{Opt}\\
&& \Sigma\doteq \left(\prod_{i\in\bar{\cal V}_v} \Sigma_i \right) \cap \left(\prod_{\beta\in\bar{\cal V}_c}\Sigma_\beta \right)
\label{Opt}
\end{eqnarray}
defined over the bipartite graph, $\bar{\cal G}\doteq (\bar{\cal V}_v,(\bar{\cal V}_f\cup\bar{\cal V}_c) ,\bar{\cal E})$, where $\bar{\cal V}_v$, $\bar{\cal V}_f$, $\bar{\cal V}_c$ and $\bar{\cal E}$ are the sets of variable-nodes, factor-function nodes, constrain-expressing nodes and edges connecting variable-nodes and factor-function and constrain-expressing nodes to each other.  Here in Eq.~(\ref{Opt}), the variable $x\doteq (x_i|i\in\bar{\cal V}_v)$ is a vector with components, $x_i$, labeled by $i$ - a variable-node from $\bar{\cal V}_v$, taking values from the set $\Sigma_i$, which can be discrete or continuous, e.g. taken values over reals. The function $f_\alpha(x_\alpha)$ in Eq.~(\ref{Opt}), associated with a factor $\alpha\in\bar{\cal V}_f$, is a function of $x_\alpha\doteq (x_i|i\sim \alpha)$ - vector constructed from variable-nodes connected to the factor $\alpha$ through an edge,  thus $i\in\alpha$ is a shortcut for $\forall i\in{\cal V}_v\mbox{ s.t. }(i,\alpha)\in\bar{\cal E}$. We assume that $\forall \alpha\in\bar{\cal V}_f$ factor function $f_\alpha:\Sigma_\alpha\to R^+$ maps from $\Sigma_\alpha\doteq \cup_{i\sim \alpha}\Sigma_i$ to the set of non-negative finite reals.
$\Sigma_\beta$ in Eq.~(\ref{Opt}), associated with a factor $\beta\in\bar{\cal V}_c$, is set of $x_\beta\doteq (x_i|i\sim \beta)$.

In the next Section we will describe LP-BP approach to solving Eq.~(\ref{Opt}) which will obviously apply to the Physics-Constrained network flow problem as well,  provided (a) transformation from the network graph ${\cal G}$ to the auxiliary graph $\bar{\cal G}$ is done according to Fig.~\ref{graph-trans} and explanations above; (b) $x$ variable in the general formulation (\ref{Opt}) is built by combining $\pi$,$\phi$ and $q$ variables; and (c) the constraints (\ref{Upsilon_i},\ref{Upsilon_ij}) are embedded in the description of the $\Sigma_\beta$ constraints.

\section{From Nonlinear Programming to Linear Programming - Belief Propagation (LP-BP)}
\label{sec:LP-BP}

In this Section we utilize the GM reformulation of the PCNF problems and discuss
transformation from Eq.~(\ref{Opt}) to the so-called Linear Programming - Belief Propagation (LP-BP). The transformationis done in three steps. First, in Section \ref{subsec:NP_to_LP_exact}, we restate Eq.~(\ref{Opt}) as an LP in the space of beliefs (proxies for probabilities). Second, in Section \ref{subsec:LP-BP_relax} we introduce the LP-BP relaxation. Finally, to get a tractable relaxation of LP-BP and thus of the original NP (\ref{Opt}) we introduce in Section \ref{subsec:LP-BP_int} part-LP-BP scheme based on an interval partitioning of the underlying space.

\subsection{Exact reformulation of the Nonlinear Programming as a Linear Programming in the space of beliefs}
\label{subsec:NP_to_LP_exact}

The optimization problem (\ref{Opt}) also allows reformulation as the Exact Linear Programming (ELP)
\begin{eqnarray}
\mbox{ELP}:\quad \min_{b(x)\in {\cal B}} \int\limits_{\Sigma} dx b(x) \sum_{\alpha\in \bar{\cal V}_f} f_\alpha (x_\alpha)
\label{ELP}
\end{eqnarray}
where $\int_\Sigma dx ...$ stands for integration (or summation when $\Sigma$ is discrete) in $x$ over $\Sigma$, and ${\cal B}$ is the following Exact Set (ES):
\begin{eqnarray}
{\cal B}\doteq\left(\{b(x)\}\left|\begin{array}{cc}
0\leq b(x)\leq 1,& \forall x\in\Sigma\\
\int\limits_\Sigma dx b(x)=1. &
\end{array}\right.\right)
\label{ES}
\end{eqnarray}

In general  the belief set, ${\cal B}$ 
is not tractable as the number of variables and the number of the set defining constraints are both infinite when $\Sigma$ contains a continuous subset, and the numbers (of variables and constraints) are exponential in the dimension even  when all the $\Sigma_i$ are discrete. This suggests construction of various relaxations of the ELP through constraint generation methods such as the cutting plane procedure - consisting simply in keeping only a finite subset of constraints from ${\cal B}$ thus expressed through a subset of beliefs, or moments. By construction any of the relaxations shows the following two key features:
\begin{itemize}
\item Optimum value of the relaxed optimization provides a low bound on the exact value of Eq.~(\ref{Opt}) (or, equivalently, of Eq.~(\ref{ELP})).
(We can also construct an upper bound presenting a feasible solution.)
\item If the optimum solution (argument) of the relaxed optimization Eq.~(\ref{ELP}) is integer (all beliefs are $0$ or $1$) then this is also an optimal solution of OPT=ELP.  In this (lucky) case we say that the gap is closed.
\end{itemize}

In the following we will discuss a particular relaxation strategy, called Linear-Programming - Belief Propagation, and then (very briefly) comment on the possibility of constructing adaptively and inhomogeneously over the graph a hierarchy of the Sherali-Adams type starting from LP-BP and proceeding with some extra (and more complex) beliefs added. We intend to make construction of the hierarchy adaptive, so that the choice of the higher-order constraints to add to the set of active constraints (included in the optimization) depends on the result/output of the preceding step.

\subsection{LP-BP relaxation}
\label{subsec:LP-BP_relax}

ES (\ref{ES}) can be restated as
\begin{eqnarray}
{\scriptsize
\hspace{-0.6cm} {\cal B}\doteq\left(\begin{array}{c} \{b(x),\\ b_\alpha(x_\alpha),\\ b_i(x_i)\}\end{array}\left|\begin{array}{cc}
0\leq b(x)\leq 1,& \forall x\in\Sigma\\
b_\alpha(x_\alpha)=\int\limits_{\Sigma \setminus \Sigma_\alpha} d(x\setminus x_\alpha) b(x), & \forall \alpha\in \bar{\cal V}_f\cup \bar{\cal V}_c,\ \forall x_\alpha\in\Sigma_\alpha\\
b_i(x_i)= \int\limits_{\Sigma_\alpha \setminus \Sigma_i} d(x_\alpha\setminus x_i) b_\alpha(x), & \forall i\in\bar{\cal V}_v,\ \forall \alpha\in\bar{\cal V}_f\cup \bar{\cal V}_c \mbox{ s.t. } \alpha\sim i,\ \forall x_i\in\Sigma_i\\
\int\limits_{\Sigma_i} dx_i b(x_i)=1 &  \forall i\in\bar{\cal V}_v
\end{array}\right.\right)}
\label{ES1}
\end{eqnarray}
where we simply added/defined beliefs associated with node- and factor- variables.  LP-BP relaxation of the ES, also called (Graph) Local Consistency Relaxation (LCR) in \cite{14WG}, consists in replacing the first two lines in Eq.(\ref{ES1}) by the range inequalities for beliefs associated with the factor variables
\begin{eqnarray}
&& {\cal B}_{LP-BP}\doteq\label{LP-BP_domain}\\ && {\small \left(\begin{array}{c} \{ b_\alpha(x_\alpha),\\ b_i(x_i)\}\end{array}\left|\begin{array}{cc}
0\leq b_\alpha(x_\alpha)\leq 1,& \forall \alpha\in\bar{\cal V}_f\cup \bar{\cal V}_c,\ \forall x_\alpha\in\Sigma_\alpha\\
b_i(x_i)= \int\limits_{\Sigma_\alpha \setminus \Sigma_i} d(x_\alpha\setminus x_i) b_\alpha(x), & \forall i\in\bar{\cal V}_v,\ \forall \alpha\in\bar{\cal V}_f\cup \bar{\cal V}_c \mbox{ s.t. } \alpha\sim i,\ \forall x_i\in\Sigma_i\\
\int\limits_{\Sigma_i} dx_i b(x_i)=1, &  \forall i\in\bar{\cal V}_v
\end{array}\right.\right)}
\nonumber
\end{eqnarray}
Then the relaxed version of the ELP is
\begin{eqnarray}
\mbox{LP-BP}:\quad \min_{\{b_i,b_\alpha\}\in {\cal B}_{LP-BP}} \sum_{\alpha\in\bar{\cal V}_f\cup \bar{\cal V}_c} \int_{\Sigma_\alpha} dx_\alpha b_\alpha(x_\alpha) f_\alpha (x_\alpha).
\label{LP_BP}
\end{eqnarray}
Since LP-BP is relaxation of the ELP, one generally observes a gap between the two:
\begin{eqnarray}
\mbox{LP-BP}\leq \mbox{ELP}.
\label{relax-LP-BP}
\end{eqnarray}

Two remarks are in order.
\begin{itemize}
\item We call the aforementioned LP-relaxation (in the space of probabilities/beliefs) of the optimization problem (\ref{Opt}) LP-BP following terminology and tradition of the Graphical Model and Belief Propagation community. See, e.g. \cite{08WJ}, and references therein. However,  exactly the same object was discussed even earlier in the combinatorial optimization community. See \cite{14ZWP} and references therein. According to the complementary terminology Eq.~(\ref{Opt}) describes the valued constrained satisfaction problem and Eq.~(\ref{LP_BP}) is called ``basic LP relaxation".

\item Even though the set (\ref{ES1}) 
is convex,  the LP-BP optimization (\ref{LP_BP}) 
is still not tractable (in the case of continuous alphabet) as description of the ${\cal B}_{LP-BP}$
set includes infinitely many constraints.

\item Sub-optimality of LP-BP is related to the fact that it ignores global constraints between beliefs by accounting only for explicit relations between factor/constraint beliefs and nodal beliefs. In other words, LP-BP allows us to only optimize over local beliefs.
\end{itemize}

\subsection{Tractable, Interval-Partitioned Relaxation of LP-BP}
\label{subsec:LP-BP_int}


The semi-infinite nature, and thus intractability of LP-BP in the case of interest when components of $x$ are continuous (or mixed) calls for developing tractable approximations of LP-BP. Specifically, given that LP-BP is a relaxation, i.e. an outer approximation - lower bound, of the original NP itself, we are interested in finding a tractable lower bound to LP-BP, so that it would also be a lower bound to the NP.

We suggest an approach which consists in partitioning each $\Sigma_i$, corresponding to an elementary continuous variable, into a finite number of intervals. Assume that such a partitioning $\Sigma_i=\cup_{a_i\in{\cal A}_i} \Sigma_{i;a_i}$, where ${\cal A}_i$ is a set of labels for non-overlapping intervals, is given. (Thus leaving discussion of an optimal partitioning for Section ?.) Then, one naturally defines a finite set of marginal beliefs associated with each interval of each elementary variable:
\begin{eqnarray}
\forall i \in\bar{\cal V}_v,\ \forall a_i\in {\cal A}_i:\quad b_{i;a_i}\doteq\int\limits_{\Sigma_{i;a_i}} d x_i b_i(x_i).
\label{b_i_a}
\end{eqnarray}
By construction, $b_{i;a}$, are all properly normalized,
\begin{eqnarray}
\forall i\in \bar{\cal V}_v:\quad \sum_{a\in{\cal A}_i} b_{i;a}=1.
\label{b_i_a}
\end{eqnarray}
Respective, and also properly normalized, finite-dimensional factor and constrain beliefs are defined according to
\begin{eqnarray}
&& \forall \alpha\in\bar{\cal V}_f,\ \forall a_\alpha=(a_i|i\sim \alpha):\quad b_{\alpha;a_\alpha}\doteq
\int\limits_{\prod_{i\sim\alpha}\Sigma_{i;a_i}} d x_\alpha b_\alpha(x_\alpha),
\label{b_alpha_a}\\
&& \forall \beta\in\bar{\cal V}_c,\ \forall a_\beta=(a_i|i\sim \beta):\quad b_{\beta;a_\beta}\doteq
\int\limits_{(\prod_{i\sim \beta}\Sigma_{i;a_i}) \cap \Sigma_\beta} d x_\beta b_\beta(x_\beta).
\label{b_beta_a}\\
&& \forall \alpha\in(\bar{\cal V}_f\cup \bar{\cal V}_c):\quad \sum_{a_\alpha} b_{\alpha;a_\alpha}=1.\label{normalization}
\end{eqnarray}
Marginalization relation between the interval partitioned node- and factor- or constraint- beliefs are also straightforward
\begin{eqnarray}
&& \forall i\in{\cal V}_v,\ \forall a_i\in{\cal A}_i,\ \forall \alpha\in (\bar{\cal V}_f\cup \bar{\cal V}_c),\ \mbox{ s.t. } \alpha\sim i:\quad b_{i;a_i}=\sum_{a_\alpha\setminus a_i} b_{\alpha;a_\alpha}.\label{b_a_consistency}
\end{eqnarray}
Then  we form the following interval-partitioned finite (thus tractable) belief polytope
\begin{eqnarray}
&& {\cal B}_{Int-Part-LP-BP}\doteq \left( \{b_{i;a_i},b_{\alpha;a_\alpha}\}\left|
\mbox{Eqs.~(\ref{b_a_consistency},\ref{b_i_a})}
\right.\right),\label{B_part_LP_BP}
\end{eqnarray}
which is, by construction, a relaxation (outer approximation) of the LP-BP polytope (\ref{LP-BP_domain}).

Next we introduce piece-wise-constant lower bound approximations for the factor functions, $f_\alpha(x_\alpha)$
\begin{eqnarray}
\forall \alpha\in\bar{\cal V}_f,\ \forall a_\alpha,\ \forall x_\alpha\in \Sigma_\alpha:\quad f_{\alpha;a_\alpha}\leq f_\alpha(x_\alpha).
\label{f_alpha_appr}
\end{eqnarray}
Combining Eqs.~(\ref{B_part_LP_BP}) with Eq.~(\ref{f_alpha_appr}) one constructs the following tractable (finite-dimensional) LP
\begin{eqnarray}
\mbox{Int-Part-LP-BP}:\quad \min_{\{b_{i;a_i},b_{\alpha;a_\alpha}\}\in {\cal B}_{Int-Part-LP-BP}}
\sum_{\alpha\in\bar{\cal V}_f, a_\alpha} b_{\alpha;a_\alpha} f_{\alpha;a_\alpha}.
\label{Int_Part_LP_BP}
\end{eqnarray}
which is provably an interval partitioned relaxation of LP-BP and thus of ELP and OPT, i.e.
\begin{eqnarray}
\mbox{Int-Part-LP-BP}\leq \mbox{LP-BP}\leq \mbox{ELP}=\mbox{OPT}.
\label{relax_Int_Part_LP_BP}
\end{eqnarray}

\section{Generalization of the LP-BP Relaxation and Associated Hierarchies}
\label{sec:hierarchies}

We saw in Subsection \ref{subsec:LP-BP_relax} that the Exact Linear-Program (ELP) in Eq. (\ref{ES}) can be relaxed into a simpler LP using the Linear-Programing Belief-Propagation (LP-BP) relaxation from Eq. (\ref{relax-LP-BP}). The LP-BP relaxation can be generalized and performed in a systematic way leading asymptotically to the exact result. This generalization results in a relaxation hierarchy of increasing tightness but also with an increasing computational complexity.

\subsection{LP-BP Hierarchy}
\label{subsec:LP-BP-hierarchy}

The key idea behind the LP-BP hierarchy is to relax Eq. (\ref{ES}) with a set of consistent beliefs involving group of variables of increasing size around more than one factor node and constrain node.
The LP-BP hierarchy is not unique as there are multiple ways of grouping variables nodes into ``super-nodes".
A set of super-nodes $\bar{\cal V}_S$ is a collection of subsets of variable nodes
\begin{align}
\bar{\cal V}_S \subset \left\{ \gamma \in {\cal P}(\bar{\cal V}_i) \right\},
\label{node_LP-BP_hierarchy}
\end{align}
where ${\cal P}(\cdot)$ denotes the power set of an ensemble. To be an admissible set of super-nodes $\bar{\cal V}_S$ should satisfy two conditions.
First, any subset of a super-node should also be a considered as a super-node
\begin{align}
\forall \gamma \in \bar{\cal V}_S, \quad \beta \subset \gamma  \Rightarrow \beta \in \bar{\cal V}_S.
\label{subsets_LP-BP_hierarchy}
\end{align}
Second, sets of variable nodes neighboring a factor node or a constrain node are super-nodes
\begin{align}
\forall \alpha \in (\bar{\cal V}_f\cup \bar{\cal V}_c), \quad \left\{ i \in {\cal V}_i \mid i \sim \alpha \right\} \in \bar{\cal V}_S.
\label{ground_LP-BP_hierarchy}
\end{align}
The generalized LP-BP relaxation of the constraints in Eq. (\ref{ES1}) based on the set of ``super-nodes" $\bar{\cal V}_S$ reads as follows
\begin{align}
{\cal B}_{LP-BP}(\bar{\cal V}_S)\doteq && {\small \left(\begin{array}{c} b_{\gamma}(x_{\gamma})\geq 0\end{array}\left|\begin{array}{cc}
b_\beta(x_\beta)= \int\limits_{\Sigma_\gamma \setminus \Sigma_\beta} d(x_\gamma\setminus x_\beta) b_\gamma(x_\gamma), & \ \forall \gamma, \beta \in\bar{\cal V}_S \mbox{ s.t. } \beta \subset \gamma,\\
\int\limits_{\Sigma_\gamma} dx_\gamma b_\gamma(x_\gamma)=1, &  \forall \gamma \in \bar{\cal V}_S
\end{array}\right.\right)}.
\label{Generalized-LP-BP}
\end{align}

The union of power sets of variable nodes around factor or constrain nodes is the minimal set of super-nodes
\begin{align}
\bar{\cal V}_{S_\text{min}} = \bigcup\limits_{\alpha \in (\bar{\cal V}_f\cup \bar{\cal V}_c)} {\cal P}\left( \left\{ i \in {\cal V}_i \mid i \sim \alpha \right\}\right),
\label{Min-gen-LP-BP}
\end{align}
and all possible combinations of variable nodes is the maximal set of super-nodes
\begin{align}
\bar{\cal V}_{S_\text{max}} = {\cal P}\left({\cal V}_i\right).
\label{Max-gen-LP-BP}
\end{align}
An LP-BP relaxation hierarchy consists of applying the generalized LP-BP relaxation (\ref{Generalized-LP-BP}) to an increasing collections of super-nodes
\begin{align}
\bar{\cal V}_{S_\text{min}}\subset \bar{\cal V}_{S_1} \subset \bar{\cal V}_{S_2} \subset \cdots \subset \bar{\cal V}_{S_\text{max}},
\label{Max-gen-LP-BP}
\end{align}
which result in LP-BP relaxations of increasing tightness. Note that the number of variables and constraints associated with an LP-BP relaxation is exponential in the size of the biggest super-node. The challenge in constructing an LP-BP hierarchy is to build small super-node sets that still provide an effective tightening.

The lowest level of LP-BP hierarchies is in general not equal to the LP-BP relaxation introduced in Eq. (\ref{LP-BP_domain}) and is always tighter
\begin{align}
{\cal B}_{LP-BP}(\bar{\cal V}_{S_{min}}) \subset {\cal B}_{LP-BP}.
\label{Min-LP-BP-relax}
\end{align}
However if every pair of factor or constrain nodes have at most one variable node as a common neighbor, then the two relaxations are equal. The technical reason behind this discrepancy comes from Condition (\ref{subsets_LP-BP_hierarchy}) that is needed for generalizing LP-BP to arbitrary set of variables. The super-set of variable nodes that is used to derived the plain LP-BP in Eq. (\ref{LP-BP_domain}) only contains sets of variable nodes neighboring factor or constrain nodes and singleton of one variable node.

Note that the highest level of LP-BP hierarchies is simply the exact set from Eq. (\ref{ES}) as it considers beliefs over all variables
\begin{align}
{\cal B}_{LP-BP}(\bar{\cal V}_{S_{max}})=\cal B.
\label{Max-LP-BP-relax}
\end{align}

\subsection{Relationship to other Relaxation Hierarchies}
\label{subsec:hierarchy-relations}

The LP-BP relaxation in Eq. (\ref{Generalized-LP-BP}) can be formulated for any set of super-nodes $\bar{\cal V}_{S_t}$. In particular the set of super-nodes can be oblivious to any GM structure contained in the problem. Although this is in general not a desirable property, it makes possible to establish a relationship between the LP-BP relaxation hierarchy and other known hierarchies.

Consider the set of super-node consisting of all subsets of variable nodes of size at most $t>0$
\begin{align}
\bar{\cal V}_{S_t} = \left\{ \gamma \in {\cal P}(\bar{\cal V}_i) \mid t \geq |\gamma| \right \}.
\label{SA-LP-BP-supernodes}
\end{align}

The sets (\ref{SA-LP-BP-supernodes}) do not take advantage of the graph structure but remain valid as super-node sets. The corresponding LP-BP relaxations ${\cal B}_{LP-BP}(\bar{\cal V}_{S_{t}})$ form a relaxation hierarchy for increasing $t$. This hierarchy is exact for levels $t\geq1+\omega(G)$ where $\omega(G)$ is the tree-width of the factor graph, leading potentially to a much smaller relaxation than (\ref{Max-LP-BP-relax}). (See related recent discussion of the interval partitioning and tree-width based solution of the Optimal Power Flow problem in \cite{15BM}.) When variables are binary this particular LP-BP hierarchy becomes equivalent to the Sherali-Adams hierarchy \cite{90SA}. However for variables with discrete alphabet, binary included, this LP-BP hierarchy is not comparable to the Lasserre moments hierarchy \cite{01Las} based on semi-definite matrices. Note that it can be shown that for any given level of the LP-BP hierarchy ${\cal B}_{LP-BP}(\bar{\cal V}_{S_{t}})$, there exist a level for which the Lassere moments hierarchy is tighter. For more information on the relationship between LP-BP and other hierarchies, we refer the reader to \cite{08WJ}.

\section{Exactness in Trees and Distributed Message Passing}
\label{sec:tree}

In the special case when the graph $\bar{\cal G}$ is a tree, it is well-known that the LP-BP relaxation to ELP is tight, see \cite{08WJ} and references there in. However when $\Sigma$ contains a continuous subset, we still need to discretize the continuous domains as in Section~\ref{subsec:LP-BP_int} to obtain a tractable lower bound given by the Int-Part-LP-BP. Now, the only inexactness, and hence the lower bound, arises from the error due to discretization.
\begin{eqnarray}
\mbox{Int-Part-LP-BP}\leq \mbox{LP-BP} =  \mbox{ELP}=\mbox{OPT}.
\label{exact_LP_BP_trees}
\end{eqnarray}

The tree structure can also be exploited to design a Dynamic Programming (DP) based algorithm to solve the Int-Part-LP-BP. The resulting algorithm has a complexity of $O(n)$. Following \cite{Dvijotham2016}, we present here an implementation of the DP which involves a single forward and backward sweep over the tree, and can be written in the form of the following message passing algorithm.

Let $p(j)$ denotes the parent of a node $j$, and $\mathcal{C}(i)$ denotes the set of children of a node $i$. Let $\mathcal{L}$ denote the set of leaves.\\

\noindent
\textbf{Forward Pass:} \\ \\
\emph{Initialization}
\begin{align}
\label{eq:message_passing_forward_initialization}
\forall i \in \bar{\cal V}_v \cap \mathcal{L}, \ \alpha = p(i),\ \forall a_i \in \mathcal{A}_i, \quad &\kappa_{i \rightarrow \alpha}(a_i) \leftarrow  0, \\
\forall \alpha \in \{\bar{\cal V}_f \cup \bar{\cal V}_c\} \cap \mathcal{L}, \ i = p(\alpha), \ \forall a_{i} \in \mathcal{A}_{i}, \quad &\gamma_{\alpha \rightarrow i}(a_{i}) \leftarrow f_{\alpha ; a_{\alpha}}, \\
& S_{processed} \leftarrow \mathcal{L}.
\end{align}
\emph{Forward traverse}
\begin{align}
\label{eq:message_passing_forward}
\mbox{Repeat until $S_{processed} = \bar{\cal V}$}, & \\
\mbox{choose} \quad & v \notin S_{processed}  \ \mbox{s.t.} \  \mathcal{C}(v) \subseteq S_{processed}, \\
\mbox{if} \ v = i \in \bar{\cal V}_v, \ \alpha=p(i): \quad & \forall a_i \in \mathcal{A}_i, \\
& \kappa_{i \rightarrow \alpha}(a_i) \leftarrow \sum_{\bar{\alpha} \in \mathcal{C}(i)}  \gamma_{\bar{\alpha}}(a_i), \\
\mbox{else if} \ v=\alpha \in \bar{\cal V}_f \cup \bar{\cal V}_c, \ i=p(\alpha): \quad & \forall a_i \in \mathcal{A}_i, \\
& \gamma_{\alpha \rightarrow i}(a_{i}) \leftarrow \min_{a_{\alpha} \in \mathcal{A}_{\alpha} \setminus a_i} \sum_{j \in \mathcal{C}(i)}  \kappa_{j}(a_{\alpha}(j)) + f_{\alpha}(a_{\alpha}) \label{eq:factor_minimization}
\end{align}

\noindent
\textbf{Backward Pass:} \\ \\
\emph{Initialization}
\begin{align}
r \in \bar{\cal V}_v = Root, \quad & a_{r}^* = \mbox{argmin}_{a_{r} \in \mathcal{A}_r} \sum_{\bar{\alpha} \in \mathcal{C}_r} \gamma_{\bar{\alpha}}(a_r), \\
& S_{assigned} \leftarrow \{r\}.
\end{align}
\emph{Backward Traverse}
\begin{align}
\label{eq:message_passing_backward}
\mbox{Repeat until $S_{assigned} = \bar{\cal V}$}, & \\
\mbox{choose} \quad & v \notin S_{assigned}  \ \mbox{s.t.} \  p(v) \subseteq S_{assigned}, \\
\mbox{if} \ v = i \in \bar{\cal V}_v:  \quad  & \mbox{continue}, \\
\mbox{else if} \ v=\alpha \in \bar{\cal V}_f \cup \bar{\cal V}_c, \ i=p(\alpha): \quad & a_{\alpha}^* \leftarrow \mbox{argmin}_{a_{\alpha} \in \mathcal{A}_{\alpha}:a_{\alpha}(i) = a_i^*} f_{\alpha}(a_{\alpha}) \\
& + \sum_{j \in \mathcal{C}(i)}  \kappa_{j}(a_{\alpha}(j)).
\end{align}

By using a finer partitioning, i.e., increasing the number of partitions in ${\cal A}_i$ it is possible to obtain very accurate lower bounds to the ELP. However, the computational complexity of the Int-Part-LP-BP as well as the corresponding DP increases rapidly as the
number of partitions increases. If $|\mathcal{A}_i| \sim t$, then $|\mathcal{A}_{\alpha}| \sim t^{deg(\alpha)}$, where $deg(\alpha)$ is the nodal degree of the factor $\alpha$. Observing that step \eqref{eq:factor_minimization} is essentially an exhaustive search over $t^{deg(\alpha)}$ elements, the computational time can grow quite fast for a given accuracy requirement on the lower bound.

Significant computational benefits can be obtained by reducing the size of the $\Sigma_i$ via pre-processing.
This can be accomplished using the so called \emph{Bound Tightening} technique, a well-known technique in the field of Contraint Programming (CP). We will describe the bound tightening scheme in the special case when the domains $\Sigma_i$ are intervals given by $\Sigma_i = [l_i, u_i]$. Then the Bound Tightening pre-processing aims at shrinking $\Sigma_i$ by solving the following optimization problems:
\begin{align}
\label{eq:global_tightening}
 l_i^{(t)} = \min_{x \in \Sigma} x_i, \quad u_i^{(t)} = \max_{x \in \Sigma} x_i.
\end{align}
The above program infers a tightened bound on each variable by propagating the bounds on the other variables via the constraints. However, the program in \eqref{eq:global_tightening} can be as difficult as the original ELP. Instead we suggest a local parallelizable sequential bound tightening scheme below.

Let $N(v)$ denote the set of neighbors of vertex $v$.
\begin{align}
\label{eq:bound_tightening}
\mbox{for} \ t = 1,2, \ldots, T: \qquad \qquad \qquad&\\
(l_{i}^{t+1}, u_i^{t+1}) \leftarrow \min/\max \quad & x_{i}, \\
\mbox{subject to}\quad  & \forall \alpha \in N(i) \cap \bar{\cal V}_c , \quad  x_{\alpha} \in \Sigma_{\alpha}, \label{eq:bound_tightening_NL_constraints} \\
                                                                &x_{j} \in [l_j^t, u_j^t], \ \forall j \in N(\alpha).
\end{align}
The bound tightening procedure described above produces a sequence of increasingly tighter bounds in each iteration. One can either continue the procedure until an approximate fixed point is reached, or terminate at any earlier stage when desirable tightening has been obtained. There are also various strategies one can use to solve the optimization problem \eqref{eq:bound_tightening}. For example, the constraints in \eqref{eq:bound_tightening_NL_constraints} can be replaced by a convex relaxation, and the resulting problem can be solved using a convex non-linear solver such as IPOPT \cite{}. This is still a valid bound since the convex relaxation will produce an interval that is a superset of the interval produced by solving \eqref{eq:bound_tightening} exactly.
Alternatively, if the number of constraint nodes in $N(i) \cap \bar{\cal V}_c$ is small (even though $N(i) \cap \{\bar{\cal V}_c \cup \bar{\cal V}_f\}$ may be large), then \eqref{eq:bound_tightening} can be solved by discretization similar to Int-Part-LP-BP, followd by exhaustive enumeration.

The combination of bound tightening and DP was shown to be very successful to solve the Optimal Power Flow problem in power distribution networks which are naturally tree structured \cite{Dvijotham2016}. Although the DP algorithm does not directly generalize to loopy graphs, the bound tightening scheme in \eqref{eq:bound_tightening} can still be utilized.

\section{Conclusions and Path Forward}
\label{sec:conclusions}

In this paper we have described ways to represent optimization and inference problems in physical flow networks as Graphical Models.  Then, focusing on the optimization (Maximum Likelihood) problems we discuss LP-BP relaxation of the resulting GM and related hierarchies. We also discuss the case when the underlying graph of relations is a tree, when LP-BP becomes exact and can also be resolved via a distributed message passing algorithm of the Dynamic Programming  type.

Even though we believe that the GM approach will help in a future to build efficient and accurate algorithmic solution of various physical flow problems, the results reviewed and presented in this manuscript are clearly preliminary.

We conclude with an incomplete list of future directions extending the material presented above.
\begin{itemize}
\item LP-BP provides a provable low-bound.  However the resulting gap may be significant.  A valuable input may be received by describing classes of physical flow problems solvable by LP-BP exactly. It is known from early works of Schlesinger \cite{76Shl}, see also \cite{07Wer,08Wer,10Wer,15Bach}, that LP-BP is exact when factors are sub-modular. The class of problems solvable exactly by LP-BP extends to the so-called symmetric fractional polymorphism class \cite{13KTZ}. On the other hand many simple (not constrained by physical potentials) network flow problems are known to be (or conjectured to be) LP-BP-gap-less too. (See e.g. \cite{10GSW} for related discussions of the message-passing approach to solving min-cost network flow problem.) It will be important to extend this line work to (a)  characterize  physical-flow GM problems which are gapless; and (b) develop an approach which allows to quantify the gap associated with LP-BP of the difficult physical flow GM formulations.

\item The fact that LP-BP provides a provable low-bound is powerful. However the bound does not extend to the more challenging case of statistical inference when LP-BP optimization is substituted by generally non-convex (due to an added entropy term) minimization of the so-called Bethe Free energy functional \cite{05YFW}. The Bethe-free energy approach is exact for GM stated for trees (then equivalent to DP), but generally it  provides neither low- no upper- bounds on marginal probabilities (or equivalently, on the corresponding normalization factors, called partition functions). It would be important to extend bounding techniques based on GM to the physical flow GM inference problem. Approximating the entropy terms via a chain rule stated solely in terms of the marginal beliefs \cite{16Ris} may be an interesting towards resolving the problem.

\item One significant advantage of LP-BP over LP of a general position is related to an expectation that it can be solved efficiently via a distributed message-passing algorithm. However, designing such provably convergent and sufficiently fast algorithm is not an easy task, which was completed for only a handful of loopy GM, noticeably for Gaussian GM under conditions of walk-summability \cite{06MJW} and matching GM \cite{15APCS}. Such distributed, efficient and provably convergent message passing algorithms are yet to be developed for the physical flow GMs.

\item If LP-BP is not optimal, it is natural to consider correcting it taking into account the non-integer part of the solution,  which is known \cite{08Joh} to have a support within a loop of the graph. Once the loopy structure is identified one may want to modify GM, or equivalently introduce some additional constraint between beliefs associated with the loops and not linked before in the bare LP-BP.  This scheme was developed in \cite{11KJC,13KJC} based on the notion of frustrated cycles and an associated Constrained Satisfaction Problem (CSP) stated in terms of beliefs optimal for the original LP-BP. Similar but different heuristic approaches were also discussed in \cite{SonJaa_nips08,SontagEtAl_uai08,SontagChoeLi_uai12} for a aGM of a general position. Such an approach, which can be viewed as an adaptive and graph-related next step (after LP-BP) in the Sherali-Adams hierarchy, was not yet discussed/tested on examples of the physical flow GMs.

\item As discussed above in Section \ref{subsec:LP-BP_int} interval partitioning is an important step in making LP-BP for GM with continuous valued variables tractable. Taking advantage of the constrained programming approach to condition variables and then partitioning the intervals adaptively constitute a promising method already tested  in \cite{Dvijotham2016} on mixed physical flow GM problems over tree graphs.  Extending this method to physical flow GM problems over loopy graphs will be our next natural step/challenge along this line of research. Notice also that finite dimensional parametrization, e.g. via mixture models \cite{mix-model}, constitute another promising alternative (to interval partitioning) for solving the continuous valued physical flow GM problems.

\item The GM-based approach (which we have just started to develop) needs to be compared to more other approaches.  In the context of the Optimum Power Flow optimization (which is bar far the most well studied PCNF optimization problem) we plan a detailed future comparison of the ``GM-based LP-BP and beyond" approach with many new results derived most recently via diverse set of SDP relaxations and related \cite{12LL,11LMBD,12Jab,13Bie,14Low1,14Low2,14MH,15MSL}.

\item It will be important to extend the GM approach to more complex PCNF problems. Of a particular interest are extensions allowing to solve PCNF of stochastic and optimization type, e.g. stated in the so-called Chance-Constrained (CC) format \cite{14BCH,16RMCA}, and also problems involving interaction of different energy system stated in terms two (or more) coupled PCNF problems, such as coordinated scheduling for interdependent electric power and natural gas infrastructures discussed in \cite{17ZRBCA}.

\end{itemize}

The authors are grateful to M. Lubin, N. Ruozzi and J.-B. Lesserre for fruitful discussions and valuable comments. The work at LANL was carried out under the auspices of the National Nuclear Security Administration of the U.S. Department of Energy under Contract No. DE-AC52-06NA25396.

\bibliographystyle{spmpsci}
\bibliography{NetFlow,matchings}

\end{document}